\documentstyle[twoside]{article}
\catcode`\@=11
\long\def\@makefntext#1{
\protect\noindent \hbox to 3.2pt {\hskip-.9pt  
$^{{\eightrm\@thefnmark}}$\hfil}#1\hfill}		

\def\thefootnote{\fnsymbol{footnote}}
\def\@makefnmark{\hbox to 0pt{$^{\@thefnmark}$\hss}}	
	
\def\ps@myheadings{\let\@mkboth\@gobbletwo
\def\@oddhead{\hbox{}
\rightmark\hfil\eightrm\thepage}   
\def\@oddfoot{}\def\@evenhead{\eightrm\thepage\hfill
\leftmark\hbox{}}\def\@evenfoot{}
\def\sectionmark##1{}\def\subsectionmark##1{}}

\oddsidemargin=\evensidemargin
\renewcommand{\thefootnote}{\fnsymbol{footnote}}

\newcounter{sectionc}\newcounter{subsectionc}\newcounter{subsubsectionc}
\renewcommand{\section}[1] {\vspace{12pt}\addtocounter{sectionc}{1} 
\setcounter{subsectionc}{0}\setcounter{subsubsectionc}{0}\noindent 
	{\tenbf\thesectionc. #1}\par\vspace{5pt}}
\renewcommand{\subsection}[1] {\vspace{12pt}\addtocounter{subsectionc}{1} 
	\setcounter{subsubsectionc}{0}\noindent 
	{\bf\thesectionc.\thesubsectionc. {\kern1pt \bfit #1}}\par\vspace{5pt}}
\renewcommand{\subsubsection}[1] {\vspace{12pt}\addtocounter{subsubsectionc}{1}
	\noindent{\tenrm\thesectionc.\thesubsectionc.\thesubsubsectionc.
	{\kern1pt \tenit #1}}\par\vspace{5pt}}
\newcommand{\nonumsection}[1] {\vspace{12pt}\noindent{\tenbf #1}
	\par\vspace{5pt}}

\newcounter{appendixc}
\newcounter{subappendixc}[appendixc]
\newcounter{subsubappendixc}[subappendixc]
\renewcommand{\thesubappendixc}{\Alph{appendixc}.\arabic{subappendixc}}
\renewcommand{\thesubsubappendixc}
	{\Alph{appendixc}.\arabic{subappendixc}.\arabic{subsubappendixc}}

\renewcommand{\appendix}[1] {\vspace{12pt}
        \refstepcounter{appendixc}
        \setcounter{figure}{0}
        \setcounter{table}{0}
        \setcounter{lemma}{0}
        \setcounter{theorem}{0}
        \setcounter{corollary}{0}
        \setcounter{definition}{0}
        \setcounter{equation}{0}
        \renewcommand{\thefigure}{\Alph{appendixc}.\arabic{figure}}
        \renewcommand{\thetable}{\Alph{appendixc}.\arabic{table}}
        \renewcommand{\theappendixc}{\Alph{appendixc}}
        \renewcommand{\thelemma}{\Alph{appendixc}.\arabic{lemma}}
        \renewcommand{\thetheorem}{\Alph{appendixc}.\arabic{theorem}}
        \renewcommand{\thedefinition}{\Alph{appendixc}.\arabic{definition}}
        \renewcommand{\thecorollary}{\Alph{appendixc}.\arabic{corollary}}
        \renewcommand{\theequation}{\Alph{appendixc}.\arabic{equation}}
        \noindent{\tenbf Appendix \theappendixc #1}\par\vspace{5pt}}
\newcommand{\subappendix}[1] {\vspace{12pt}
        \refstepcounter{subappendixc}
        \noindent{\bf Appendix \thesubappendixc. {\kern1pt \bfit #1}}
	\par\vspace{5pt}}
\newcommand{\subsubappendix}[1] {\vspace{12pt}
        \refstepcounter{subsubappendixc}
        \noindent{\rm Appendix \thesubsubappendixc. {\kern1pt \tenit #1}}
	\par\vspace{5pt}}

\newcommand{\textlineskip}{\baselineskip=13pt}
\newcommand{\smalllineskip}{\baselineskip=10pt}

\def\eightcirc{
\begin{picture}(0,0)
\put(4.4,1.8){\circle{6.5}}
\end{picture}}
\def\eightcopyright{\eightcirc\kern2.7pt\hbox{\eightrm c}} 

\newcommand{\copyrightheading}[1]
	{\vspace*{-2.5cm}\smalllineskip{\flushleft
	{\footnotesize International Journal of Modern Physics A, #1}\\
	{\footnotesize $\eightcopyright$\, World Scientific Publishing
	 Company}\\
	 }}

\def\abstracts#1#2#3{{
	\centering{\begin{minipage}{4.5in}\baselineskip=10pt\footnotesize
	\parindent=0pt #1\par 
	\parindent=15pt #2\par
	\parindent=15pt #3
	\end{minipage}}\par}} 

\def\keywords#1{{
	\centering{\begin{minipage}{4.5in}\baselineskip=10pt\footnotesize
	{\footnotesize\it Keywords}\/: #1
	 \end{minipage}}\par}}

\renewenvironment{thebibliography}[1]
	{\frenchspacing
	 \ninerm\baselineskip=11pt
	 \begin{list}{\arabic{enumi}.}
	{\usecounter{enumi}\setlength{\parsep}{0pt}
	 \setlength{\leftmargin 12.7pt}{\rightmargin 0pt} 
	 \setlength{\itemsep}{0pt} \settowidth
	{\labelwidth}{#1.}\sloppy}}{\end{list}}

\newcounter{itemlistc}
\newcounter{romanlistc}
\newcounter{alphlistc}
\newcounter{arabiclistc}

\newcommand{\fcaption}[1]{
        \refstepcounter{figure}
        \setbox\@tempboxa = \hbox{\footnotesize Fig.~\thefigure. #1}
        \ifdim \wd\@tempboxa > 5in
           {\begin{center}
        \parbox{5in}{\footnotesize\smalllineskip Fig.~\thefigure. #1}
            \end{center}}
        \else
             {\begin{center}
             {\footnotesize Fig.~\thefigure. #1}
              \end{center}}
        \fi}

\newcommand{\tcaption}[1]{
        \refstepcounter{table}
        \setbox\@tempboxa = \hbox{\footnotesize Table~\thetable. #1}
        \ifdim \wd\@tempboxa > 5in
           {\begin{center}
        \parbox{5in}{\footnotesize\smalllineskip Table~\thetable. #1}
            \end{center}}
        \else
             {\begin{center}
             {\footnotesize Table~\thetable. #1}
              \end{center}}
        \fi}

\def\@citex[#1]#2{\if@filesw\immediate\write\@auxout
	{\string\citation{#2}}\fi
\def\@citea{}\@cite{\@for\@citeb:=#2\do
	{\@citea\def\@citea{,}\@ifundefined
	{b@\@citeb}{{\bf ?}\@warning
	{Citation `\@citeb' on page \thepage \space undefined}}
	{\csname b@\@citeb\endcsname}}}{#1}}

\newif\if@cghi
\def\cite{\@cghitrue\@ifnextchar [{\@tempswatrue
	\@citex}{\@tempswafalse\@citex[]}}
\def\citelow{\@cghifalse\@ifnextchar [{\@tempswatrue
	\@citex}{\@tempswafalse\@citex[]}}
\def\@cite#1#2{{$\null^{#1}$\if@tempswa\typeout
	{IJCGA warning: optional citation argument 
	ignored: `#2'} \fi}}

\def\pmb#1{\setbox0=\hbox{#1}
	\kern-.025em\copy0\kern-\wd0
	\kern.05em\copy0\kern-\wd0
	\kern-.025em\raise.0433em\box0}

\def\fnt#1#2{\footnotetext{\kern-.3em
	{$^{\mbox{\scriptsize #1}}$}{#2}}}

\def\fpage#1{\begingroup
\voffset=.3in
\thispagestyle{empty}\begin{table}[b]\centerline{\footnotesize #1}
	\end{table}\endgroup}

\def\runninghead#1#2{\pagestyle{myheadings}
\markboth{{\protect\footnotesize\it{\quad #1}}\hfill}
{\hfill{\protect\footnotesize\it{#2\quad}}}}
\font\tenrm=cmr10
\font\tenit=cmti10 
\font\tenbf=cmbx10
\font\bfit=cmbxti10 at 10pt
\font\ninerm=cmr9

\font\eightrm=cmr8

\font\sevenrm=cmr7
\font\fiverm=cmr5

\def\qed{\hbox{${\vcenter{\vbox{			
   \hrule height 0.4pt\hbox{\vrule width 0.4pt height 6pt
   \kern5pt\vrule width 0.4pt}\hrule height 0.4pt}}}$}}

\renewcommand{\thefootnote}{\fnsymbol{footnote}}	

\begin{document}

\runninghead{H.-P. Pavel et al.} 
{Squeezed gluon condensate and quark confinement $\ldots$}

\normalsize\textlineskip
\thispagestyle{empty}
\setcounter{page}{1}

\copyrightheading{}			

\vspace*{0.88truein}

\fpage{1}
\centerline{\bf SQUEEZED GLUON CONDENSATE AND QUARK CONFINEMENT}
\vspace*{0.035truein}
\centerline{\bf IN THE GLOBAL COLOR MODEL OF QCD}
\vspace*{0.37truein}
\centerline{\footnotesize H.-P. PAVEL\footnote{Electronic 
address: pavel@darss.mpg.uni-rostock.de}, D. BLASCHKE, 
G. R\"OPKE}
\vspace*{0.015truein}
\centerline{\footnotesize\it Fachbereich Physik, Universit\"at Rostock, 
Universit\"atsplatz 1}
\baselineskip=10pt
\centerline{\footnotesize\it D-18051 Rostock, Germany}
\vspace*{10pt}
\centerline{\footnotesize V.N. PERVUSHIN}
\vspace*{0.015truein}
\centerline{\footnotesize\it Bogoliubov Laboratory for Theoretical Physics, 
Joint Institute for Nuclear Research}
\baselineskip=10pt
\centerline{\footnotesize\it 141980 Dubna, Russia}
\vspace*{0.225truein}

\vspace*{0.21truein}
\abstracts{
We discuss how the presence of a squeezed gluon vacuum might lead to quark
confinement in the framework of the global colour model of QCD. 
Using reduced phase space quantization of massive vector theory
we construct a Lorentz invariant and colourless squeezed gluon
condensate and show that it induces a permanent, nonlocal quark
interaction (delta-function in 4-momentum space),
which according to Munczek and Nemirovsky might lead to quark confinement.
Our approach makes it possible to relate the strength of this effective
confining quark interaction to the strength of the physical gluon condensate. 
}{}{}


\pagebreak

\textheight=7.8truein
\setcounter{footnote}{0}
\renewcommand{\thefootnote}{\alph{footnote}}

\section{Introduction}
\noindent
Phenomenological models have been developed for the low energy sector
of QCD which are rather successful in describing hadrons and their
properties. Here the aspect of chiral symmetry is understood much better
than that of quark and gluon confinement.
The mechanism for dynamical chiral symmetry breaking is identified fairly
well since the work by Nambu and Jona-Lasinio (NJL)\cite{nambu}. For recent
reviews on the application of the NJL-model to low energy QCD, see \cite{njl}.
For the explanation of confinement many different models have been
discussed. We will refer in the present work to approaches based
on the global colour model \cite{mn,cahill,rw,tandy} with nonlocal confining
quark interactions. These approaches address quark and gluon confinement 
via the criterion of absence of real $q^2$ poles for the propagators 
\cite{pag}, for a review see \cite{rw}.
Such confining quark models use effective gluon propagators,
which have infrared singularities like $1/k^4$ or a
delta function $\delta(k)$ at low energy \cite{pag,mn,ebert,fr}.
These phenomenological approaches have proven successful in explaining low
energy hadronic observables \cite{fr,tandy} and in adressing the problem of
chiral vs. deconfinement transition at finite temperatures \cite{BBKR}.
However, the question for the mechanism which leads
to the infrared singularities of the gluon propagator remains open.
There is evidence from detailed studies \cite{bbz,bp} of the
Schwinger-Dyson equations of QCD for a strong infrared
enhancement of the gluon propagator due to the non-Abelian character 
of the theory and in particular due to the gluon-gluon self coupling.
On the other hand there are several approaches
which relate the confinement problem to the question of the
true physical vacuum of QCD.

In more rigorous treatments of QCD, the problem of the vacuum is
well known.
For instance, the simple perturbative vacuum is unstable \cite{sav},
and there is no stable (gauge invariant) coherent vacuum in Minkowski space
\cite{leut}.
Therefore, in the context of the construction of a gauge invariant,
stable QCD vacuum in Minkowski space, the squeezed condensate of gluons has
become a topic of great interest \cite{celenza} - \cite{DB}.
From the physical point of view the squeezed state differs from the coherent
one by the condensation of colour singlet gluon pairs rather than
condensation of single gluons.
It has been discussed before that a squeezed vacuum in form of Gaussian
fields \cite{dosch,biro1,kogan} leads to confinement of
quarks via the criterion of the linearly rising static
potential \cite{biro1} and via  the area law behaviour of the Wilson loop
\cite{kogan}.

The present paper is devoted to the study of confinement as a possible
consequence of the squeezed vacuum.
Our approach differs from the previous ones addressing the squeezed vacuum
\cite{biro1,mishra1,kogan} in the following ways.
We will not try to generate the squeezed vacuum by treating selfinteractions
of gluons but we will use the squeezed condensate as a semi-phenomenological
input.
In difference to e.g. \cite{kogan}, where all gauge degrees of freedom
are squeezed and only afterwards a projection to the gauge invariant
sector is performed, we shall apply the squeezing transformation only on
the physical degrees of freedom.  
Furthermore, in contrast to previous work we discuss confinement by using the
criterion of the absence of poles for the propagators \cite{pag}
as used in the approaches based on the global colour model \cite{mn}. 
We will prepare the squeezed vacuum by
macroscopically populating it with zero momentum gluon pairs and study the
changes in the analytical properties of the propagators caused by the change
of the vacuum structure. This procedure parallels the original idea by
Bogoliubov \cite{NN} for the explanation of the Landau sound in a superfluid
liquid according to which the change in the excitation spectrum of the theory
at low energies is due to the condensation of a macroscopic number
of particles in the zero momentum state.
As a first step we shall assume, inspired by the success of the approaches 
based on the global colour model,
that the gluon selfinteractions in the low energy region only lead to
the formation of a squeezed gluon condensate and include it into an 
otherwise Abelian approximation to QCD. For simplicity we shall ignore
here the effects of the gluon selfinteractions in the high energy region
leading to asymptotic freedom.
In order to control the infrared divergence of massless gluons 
we start in a large finite volume and give the 
gluons a small mass inversely proportional to the volume.
Finally we take the infinite volume limit which leads to a Lorentz invariant 
squeezed condensate of massless gluons and at the same time to a 
4-momentum $\delta$-function interaction between the quarks.
A type of delta function interaction has been considered for the computation
of the meson spectra on the level of Schwinger-Dyson and Bethe-Salpeter
equations in both Minkowski space \cite{pag,mn,ebert} and Euclidean space
\cite{pag,rw,mn,fr}. Munczek and Nemirovsky \cite{mn}
have shown that such a delta function interaction removes the poles of the
effective quark propagator constructed by using the Schwinger-Dyson equation.
The corresponding Bethe-Salpeter equation on the other hand
has bound state solutions.

A well-defined massless limit is not straightforward because the
covariant propagator of massive vector bosons has a singularity at mass $M=0$.
Although the singular term drops out when the massive vector field is
coupled to a conserved current as in QED, for the
explicit construction of the squeezed vacuum it is necessary to have
a physical representation of the vector field such that it smoothly turns 
into the gauge field in the massless limit. 
We shall use the results of \cite{pav}
where it has been shown that massive vector theory can be quantized in such 
a way that its massless limit agrees with the theory of photons.
In contrast to the approach by St\"uckelberg \cite{stueckel} where
ghost fields are introduced in order to maintain manifest Lorentz covariance,
in \cite{pav} a massive QED theory with a good massless 
limit is obtained in the framework of reduced phase space quantization 
\cite{HeP} without introducing any additional
unphysical degrees of freedom, but by choosing
certain nonlocal dynamical degrees of freedom. 

The present paper is organized as follows: In Section 2 we briefly review
how quark confinement in the framework of the global colour model 
has been discussed by Munczek and Nemirovsky. 
In Section 3 the squeezed condensate of gluons is constructed.
In Section 4 we derive the effective quark Lagrangian in the 
presence of the squeezed condensate. 
We show that the effective quark interaction induced by the squeezed 
condensate is a momentum delta function and relate
the strength of the $\delta$- function interaction to the value
of the physical gluon condensate.
Section 5 finally contains our conclusions.  

\section{Quark confinement in the global colour model of QCD}
\label{confine}
\noindent
The global colour model \cite{mn,cahill} is a dynamical quark model with
an effective current-current interaction, which includes highly 
nonperturbative gluon dynamics. 
It can describe many features of hadronic physics quite successfully
\cite{tandy}.
In order to briefly review how it results from full QCD,
we start with the QCD generating functional $Z_{QCD}$ for Green functions in 
the absence of external sources
\begin{equation}
Z_{QCD} = {1\over N} \int DAD\Psi D\bar{\Psi}
\exp\left[i\int d^4x\left({\cal L}_\Psi+{\cal L}_A+{\cal L}_{A\Psi} 
\right)\right]~.
\end{equation}
Here
\begin{equation}
{\cal L}_\Psi = \bar{\Psi}(x)\left(i\not\!\partial-\hat{m}\right)\Psi(x)
\end{equation}
is the quark Lagrangian with  the diagonal current mass matrix $\hat{m}$ and
\begin{equation}
{\cal L}_{A} = -{1\over 4} G^{a\mu\nu} G_{\mu\nu}^{a}
\end{equation}
is the Yang-Mills Lagrangian with the non-Abelian field strength tensor
$G_{\mu\nu}^a\equiv \partial_{\mu}A_\nu^a-\partial_{\nu}A_\mu^a 
+gf^{abc}A^b_{\mu}A^c_{\nu}$.
We have not included
gauge-fixing terms and associated quantities explicitly since they are not 
needed for the purpose of our discussion.
The quark currents 
$J^{\mu}_a\equiv g\bar\Psi\gamma_\mu {\lambda^a\over 2}\Psi$  are
minimally coupled to the gluon fields
\begin{equation}
{\cal L}_{A\Psi}=A^a_{\mu}J^{\mu}_a~.
\end{equation}
After integration over the gluon field $Z_{QCD}$ takes the form
\begin{equation}
\label{ZPsi}
Z_{QCD} = {1\over N'} \int D\psi D\bar{\psi}
\exp\big[i\int d^4x{\cal L}_\Psi +iW(J_a^{\mu})\big]~,
\end{equation}
where the gluon action functional
\begin{equation}
W(J_a^{\nu})=\sum_{m=2}{1\over m!}\int dx_1 dx_2.. dx_m
G_{\nu_1...\nu_m}^{a_1....a_m}(x_1,....,x_m)
J_{a_1}^{\nu_1}(x_1)...J_{a_m}^{\nu_m}(x_m)
\end{equation}
is written as a functional Taylor expansion in terms of the quark currents 
thus defining
$G_{\nu_1...\nu_m}^{a_1....a_m}(x_1,....,x_m)$ as the connected $m$-point
Green functions for the gluon field.\footnote{The term with $m=1$ does not 
occur because of the colour neutrality of the vacuum.}
In the global colour model the expansion of $W(J_a^{\mu})$ is truncated
at the level of the gluon two-point function $G_{\mu\nu}^{ab}(x-y)$.
Thus the effective current-current coupling in the global colour model is
\begin{equation}
 W_{GCM}[J_a^{\nu}]=
 -{1\over 2}\int d^4x\int d^4y J_a^{\mu}(x) 
G^{ab}_{\mu\nu}(x-y) J_b^{\nu}(y)~,
\end{equation}
which represents a four-fermion interaction term.
In the Feynman gauge the gluon propagator simplifies to
\begin{equation}
G_{\mu\nu}^{ab}(x-y)=-i\delta^{ab}g_{\mu\nu}G(x-y).
\end{equation}
Note that for QED the truncation at the level of the two-point function
is exact and $G(x-y)$ corresponds to perturbative one-photon exchange.
In QCD, on the other hand, the two-point function contains highly 
nonperturbative contributions due to the gluon selfinteractions,
such as the influence of a gluon condensate in the low energy region and
asymptotic freedom in the high momentum limit.

At present it is impossible yet to calculate the nonperturbative
gluon propagator from QCD. In the global colour model, several ans\"atze
have been used \cite{rw} for the nonperturbative gluon propagator.
The simplest one, which may lead to quark confinement in the sense
of absence of poles on the real $q^2$ axis in the quark propagator
has been introduced by Munczek and Nemirovsky \cite{mn} in the form
\begin{equation}
i{4\over 3}g^2{G(q)\over (2\pi)^4}=-{1\over4}\eta^2\delta^4(q)-U(q)
\label{deltaU}
\end{equation}
Using just the $\delta$- function part of the interaction and the rainbow
approximation for the vertex function, the Schwinger--Dyson
equation for the quark self energy
\begin{equation}
\label{sde}
i \Sigma(q)= {4\over 3}{g^2} \int {d^4k\over (2\pi)^4} G(k-q)
\gamma^{\mu} S(k)\gamma_{\mu}~,
\end{equation}
reduces to an algebraic equation. The solution for the inverse dressed
quark propagator is given by
\begin{equation}
iS^{-1}(q)\equiv \not{\! q}-m-\Sigma(q)\equiv \not{\! q} A(q) + B(q)~,
\end{equation}
where the solution functions $A(q)$ and $B(q)$ have been found by
Muncek and Nemirovsky \cite{mn} to conspire in such a way
that $S^{-1}(q)$ has no zeros for real $q^2>0$,
which according to Refs. \cite{pag,mn} is signalling confinement.
The coefficient $\eta^2=1.14~{\rm GeV}^2$ of the 
$\delta$-function ansatz (\ref{deltaU}) has been fixed in \cite{mn}
by the rho meson mass value obtained from the solution of the
corresponding Bethe-Salpeter equation in the ladder 
approximation\footnote{
The ladder approximation for the BSE is consistent with the 
rainbow approximation for the Schwinger-Dyson equation (\ref{sde}).
In Ref. \protect{\cite{BSR}} a consistent approximation scheme beyond
the rainbow - ladder approximation has been developed.}.

A form of the residual interaction $U(q)$ which reproduces the meson decay 
constants and provides asymptotic freedom has been given by \cite{fr}.
However it is not yet known, which approximations have to be made in
calculating the effective gluon action $W[J]$ of QCD such that a successful
phenomenological model with confinement and asymptotic freedom like in
Ref.\cite{fr} can be derived.

We shall show in the present paper that it is possible to obtain
the delta function interaction
under the assumption that a squeezed condensate of zero-momentum gluon pairs
is present in the QCD vacuum.

\section{Construction of a squeezed gluon vacuum}
\label{squeeze}
\noindent
It is a generally believed that the gluon selfinteractions in QCD lead to 
highly nontrivial infrared effects such as physical gluon condensation.

The gluon condensate is defined as
the expectation value of the local gluonic operator 
$\alpha_s G_{\mu\nu}^{a}(x)G^{\mu\nu a}(x)$ 
in the nonperturbative QCD vacuum \cite{SVZ}
\begin{eqnarray}
\label{glc}
\langle{}\alpha_sG^2\rangle ~
&\equiv & \langle{}\alpha_sG_{\mu\nu}^{a}(0)G^{\mu\nu a}(0)\rangle ~.
\end{eqnarray}

Approximate empirical values for the physical gluon condensate
are the estimate
$\langle{}\alpha_s G^2\rangle ~\simeq 0.04~ {\rm GeV^4}$ 
by Shifman, Vainshtein and Zakharov \cite{SVZ}
and the update average value 
$\langle{}\alpha_s G^2\rangle ~=(0.071\pm 0.009)~ {\rm GeV^4}$ 
obtained by Narison \cite{Narison} in a recent
analysis of heavy quarkonia mass-splittings in QCD.

In the following we shall construct a squeezed gluon vacuum in an explicit way
as a possible approximate model for the true QCD vacuum.
The infrared singularity of the gluons as well as the
their selfinteraction are generally believed to be the origin of the existence
of a physical gluon condensate in the QCD vacuum.
In phenomenological approaches based on the global colour model, 
where full QCD is truncated at the level of the gluon two-point function, 
which is correct only for Abelian theories, 
the gluon selfinteractions are included in the low energy region via
some confinement term and in the high energy region via
a term leading to asymptotic freedom. Inspired by their success to explain
many features of hadronic physics we shall assume as a first step
that the only effect of the gluon selfinteraction 
is to ensure the existence and stability of the gluon condensate,
and shall consider an otherwise Abelian approximation to QCD.
For simplicity we shall also neglect the high energy effects  due to
the gluon selfcoupling such as asymptotic freedom.
In order to isolate the infrared singular zero momentum component of the 
gluon field
from the nonzero ones we put the theory into a large but finite volume $V$.
Furthermore, in order to regularize the infrared
singularity we shall give the gluons a small mass.  
Hence we shall start with the simpler theory of massive Abelian QCD 
in a large finite volume and shall show that it is possible
to include a squeezed
gluon condensate of zero momentum gluon pairs in the infinite volume limit
by explicit construction.
For this we shall let the small mass vanish like $1/V$ in the large volume 
limit after quantization of the theory. By calculating the value of
the gluon condensate in the constructed squeezed vacuum in the 
zero momentum approximation we shall then find an estimate for the
squeezing strength due to the selfinteraction of the gluons. 

\subsection{Construction of a squeezed gluon vacuum in Abelian QCD}
\noindent
For the construction of the squeezed vacuum we shall first consider the 
Lagrangian of massive Abelian QCD given by
\begin{equation}
{\cal L}_{\rm Abel}[M,V_\mu,\Psi]=
-{1\over 4}F_{\mu\nu}^aF^{a\mu\nu}+{1\over 2}M^2V_\mu^{a2}+
\bar\Psi(i\not{\!\partial}-m)\Psi-J^{\mu}_aV_{\mu}^a~, 
\label{LAbel}
\end{equation}
with the Abelian field strengths 
$F^a_{\mu\nu}=\partial_{\mu}V^a_{\nu}-\partial_{\nu}V^a_{\mu}$.
In order to construct the squeezed vacuum as described above,
it is necessary to quantize massive Abelian 
QCD in such a way that it has a good massless limit.
Following the work of \cite{pav} we shall use
reduced phase space quantization, where this can be achieved without the 
introduction of additional unphysical ghost degrees of freedom.

\subsubsection{Reduction to physical variables}
\noindent
The Euler-Lagrange equation correponding to $V^a_0$ is not an
equation of motion but a constraint. $V^a_0$ is therefore not a dynamical
variable and has to be eliminated using the constraint equation before
the remaining reduced phase space can be quantized. 
As shown in \cite{pav} and briefly reviewed in the Appendix, the reduced
Lagrangian obtained from (\ref{LAbel}) by eliminating $V_0$ can be expressed
in terms of new nonlocal variables $\tilde{V}_k^a$ and $\tilde{\Psi}$ as
\begin{eqnarray}
{\cal L}^{\rm red}_{\rm Abel}[\tilde{V}_k^a,\tilde{\Psi}] &=& 
{1\over 2}\left(\dot{\tilde{V}}_i^a P^{-1}_{ij}
\dot{\tilde{V}}_j^a+
          \tilde{V}_i^a(\vec{\partial}^2-M^2)P^{-1}_{ij}\tilde{V}_j^a\right)
            \nonumber\\
               & & -J_i^a\tilde{V}_i^a
                   +{1\over 2}J_0^a{1\over \vec{\partial}^2-M^2}J_0^a 
                  +\bar{\tilde{\Psi}}(i\not{\!\partial}-m)\tilde{\Psi}~.
\label{redL}
\end{eqnarray}
The nonlocal fields $\tilde{V}_k^a$ are defined as
\begin{equation}
\tilde{V}_k^a=P_{kj}V_j^a
\end{equation}
with $P_{ij}$ given by
\begin{equation}
P_{ij}\equiv
\delta_{kj}-{\partial_k\partial_j\over \vec{\partial}^2-M^2}=
\delta_{ij}^T-{M^2\over \vec{\partial}^2-M^2}\delta_{ij}^{||}
\label{Projop}
\end{equation}
where we have used the longitudinal and transverse projection operators
$\delta_{ij}^{||}\equiv \partial_i\partial_j/ \vec{\partial}^2$ and
$\delta_{ij}^T\equiv \delta_{ij}-\delta_{ij}^{||}$.
In contrast to the massless case, $P_{ij}$ is invertible and $P^2\neq P$.
The nonlocal fermionic field $\tilde{\Psi}$ is defined as
\begin{equation}
\label{psired}
\tilde{\Psi}\equiv \exp\left(ig{\lambda^a\over 2}{1\over \vec{\partial}^2-M^2}
                         \partial_iV_i^a \right)\Psi~.
\end{equation}
Since this is only a phase transformation, the corresponding
currents $J^a_{\mu}$ remain unchanged.

The nonlocal fields $\tilde{V}_i^a$ differ from the standard local ones $V_i^a$
in that their longitudinal components are shortened relative to those of
$V_i^a$. In the massless limit they are shortened to zero i.e. projected out 
and hence reduce to the corresponding transverse gauge fields. 
The zero momentum modes of the nonlocal fields on the other hand
coincide with those of the local vector field, since
$P_{ij}(\vec{q}=0)=\delta_{ij}$:
\begin{eqnarray}
\lim_{M\to 0}\left(\tilde{V}_i^a(t,\vec{p}= 0),\tilde{\Pi}_i^a(t,\vec{p}= 0)
\right)&=& 
\left({\cal A}_i^a(t),{\cal E}_i^a(t)\right)\nonumber\\
\lim_{M\to 0}\left(\tilde{V}_i^a(t,\vec{p}),\tilde{\Pi}_i^a(t,\vec{p})\right)
&=& 
\left(A^{Ta}_i(t,\vec{p}),E^{Ta}_i(t,\vec{p})\right)\ \ \ \ \ \vec{p}\neq 0
\label{M=0limit}
\end{eqnarray}
with the three spatial zero momentum components ${\cal A}_i^a$ and 
${\cal E}_i^a$ and the two transverse nonzero momentum components
$A^{Ta}_i$ and $E^{Ta}_i$
of the gauge fields and the corresponding canonically conjugate 
electric fields.
For a small mass we can regard the massive nonlocal field 
as a smooth interpolation between the three gauge invariant zero momentum
components and the two transverse components of the high momentum limit.
In this scenario one could therefore conclude that the zero momentum component 
of the Abelian gauge field has three dynamical degrees of freedom in contrast
to the non zero momentum components which have only two transverse degrees
of freedom.

In order to construct the squeezed gluon vacuum via a canonical transformation
we shall now pass to the Hamiltonian formalism.
The Hamiltonian corresponding to (\ref{redL}) is given in terms of the fields 
$\tilde{V}_i^a,\tilde{\Psi}$ and the corresponding canonical
momenta $\tilde{\Pi}_i^a,\tilde{\Pi}$ by
\begin{equation}
H_{\rm Abel}^{\rm red}=H_0+H'~,
\label{Hred}
\end{equation}
where the free and interacting parts read
\begin{eqnarray}
H_0\!\!\!\! &=&\!\!\!\! \int d^3\vec{x}\left\{{1\over 2}\left(\tilde{\Pi}_i^aP_{ij}
 \tilde{\Pi}^a_j-\tilde{V}^a_i(\vec{\partial}^2-M^2)P^{-1}_{ij}V^P_j\right)+
\tilde{\Pi}^a\gamma_0(\vec{\gamma}\cdot\vec{\partial}+m)\tilde{\Psi}
\right\}\\
H'\!\!\!\! &=&\!\!\!\! \int d^3\vec{x}
\left(\tilde{V}_iJ_i - {1\over 2}J_0{1\over \vec{\partial}^2-M^2}J_0\right)~.
\end{eqnarray}
Note that the second term in $H'$ corresponds to a Yukawa potential.

The theory is now quantized in terms of the nonlocal $\tilde{V}_i^a(x)$
and the canonical conjugate momenta $\tilde{\Pi}_i^a$ 
by imposing the canonical commutation relations
\begin{equation}
i[\tilde{\Pi}^a_k(\vec{x},t),\tilde{V}^b_j(\vec{y},t)]
=\delta^{ab}\delta_{kj}\delta(\vec{x}-\vec{y}).
\end{equation}
The free Hamiltonian is diagonalized by the interaction picture fields
\begin{eqnarray}
\tilde{v}_i^a(\vec{x},t)\!\!\! &=&\!\!\! {1\over V}\sum_{\vec{q}}
{1\over\sqrt{2\omega(\vec{q})}}\sum_{\lambda=1,2,3}
\left(a^a(\lambda,\vec{q})\tilde{\epsilon}_i(\lambda,\vec{q})
e^{-i(\omega(\vec{q})t-\vec{q}\cdot\vec{x})} + h.c.\right),\nonumber\\
\tilde{\pi}_i^a(\vec{x},t)\!\!\! &=&\!\!\! -i{1\over V}\sum_{\vec{q}}
\sqrt{{\omega(\vec{q})\over 2}}\sum_{\lambda=1,2,3}
\left(a^a(\lambda,\vec{q})\epsilon_i(\lambda,\vec{q})
e^{-i(\omega(\vec{q})t-\vec{q}\cdot\vec{x})} - h.c.\right),
\label{arepr}
\end{eqnarray}
with the creation and annihilation operators satisfying
\begin{equation}
[a_a(\lambda,\vec{q}),a_b^{+}(\lambda',\vec{q}')]=\delta_{ab}
\delta_{\lambda\lambda'}
(2\pi)^3\delta(\vec{q}-\vec{q}')~,
\end{equation}
and with the real nonlocal polarization vectors 
$\tilde{\epsilon}_i(\lambda,\vec{q})$ and $\epsilon_i(\lambda,\vec{q})$
satisfying the completeness relations
\begin{eqnarray}
\sum_{\lambda}\tilde{\epsilon}_i(\lambda,\vec{q})
\tilde{\epsilon}_j(\lambda,\vec{q})
&=&\delta_{ij}-{q_i q_j\over \vec{q}^2+M^2}~,\nonumber\\
\sum_{\lambda}\epsilon_i(\lambda,\vec{q})
\epsilon_j(\lambda,\vec{q})
&=&\delta_{ij}+{q_i q_j\over M^2}~.
\end{eqnarray}

\subsubsection{A simple model for the squeezed vacuum}
\noindent
From the representation (\ref{arepr}) we obtain the following expressions 
for the vacuum expectation values of squares of the zero momentum
components of the nonlocal vector fields 
and their conjugate momenta in the limit of small mass M
\begin{eqnarray}
\lim_{M\to 0}~\langle 0|\tilde{v}^a_i(\vec{q}=0)^2|0\rangle &=&
\langle 0|\left({\cal A}^a_i\right)^2|0\rangle=
{12\over M V}~,\\
\lim_{M\to 0}~\langle 0|\tilde{\pi}^a_i(\vec{q}=0)^2|0\rangle &=& 
\langle 0|\left({\cal E}^a_i\right)^2|0\rangle=
{12~M\over V}~,
\label{calE^2pert}
\end{eqnarray}
using the massless fields ${\cal A}^a_i$ and ${\cal E}^a_i$ according to
(\ref{M=0limit}).
Note that the vacuum expectation values of these zero momentum components
are Lorentz invariant expressions as discussed in the Appendix.
For $M=0$ the first term is singular which corresponds to
the well known infrared problem common to all massless theories.
One way to regularize the infrared singularity is to have a volume dependent 
mass $M_{\rm sq}(V)$ which vanishes in the infinite volume limit.
One particular regularizing choice is
to let  $M_{\rm sq}$ behave inversely proportional to the volume, 
\begin{equation}
\label{msq}
M_{\rm sq}=1/(2C_0 V)~,
\end{equation}
with some open parameter $C_0$, which will be fixed below.
The corresponding Fock vacuum turns into the squeezed vacuum as will be shown
in the following.
We denote the vacuum corresponding to the mass (\ref{msq}) by 
$|0_{\rm sq}\rangle$.
Using the mass dependence (\ref{msq}) in (\ref{calE^2pert}) 
one easily finds the following expectation values for the
squares of ${\cal A}^a_i$ and ${\cal E}^a_i$ in the infinite volume limit
\begin{eqnarray}
\langle 0_{\rm sq}|\left({\cal A}^a_i\right)^2|0_{\rm sq}\rangle~&=&24~C_0~,
\nonumber\\ 
\langle 0_{\rm sq}|\left({\cal E}^a_i\right)^2|0_{\rm sq}\rangle~&=&
{\cal O}(V^{-2})~.
\label{calE^2sq}
\end{eqnarray}
Such vacua are called squeezed vacua \cite{celenza}-\cite{DB}.

In order to understand the meaning of the squeezed vacuum better
it is useful to consider for comparison the case of a perturbative
vacuum $|0\rangle$ corresponding to a very small but finite and
volume independent mass $M$.
For a large finite volume the squeezed vacuum $|0_{\rm sq}\rangle$ 
can be related to 
the perturbative vacuum $|0\rangle$ by a unitary transformation 
\begin{equation}
|0_{\rm sq}\rangle=U_{\rm sq}^{-1}|0\rangle~,
\end{equation}
whose action on the zero momentum components of the fields is given by
the unitary squeezing operator
\begin{equation}
\label{sqop}
U_{\rm sq}^{(0)}=
\exp\left[i {f_0\over 2}
          \Big({\cal A}^a_i{\cal E}^a_i
                +{\cal E}^a_i{\cal A}^a_i\Big)
\right]
\end{equation}
with the squeezing parameter
\begin{equation}
f_0 = {1\over 2}\ln \left(2MC_0V\right)~.
\end{equation}
We see that the coefficient $C_0$ is a measure of the squeezing strength 
and is an open parameter to be fixed below.
The squeezing operation (\ref{sqop}) corresponds to a Bogoliubov 
transformation. Bogoliubov \cite{NN} used such a transformation
for nonzero momentum modes to rediagonalize the
Hamiltonian after having macroscopically
filled the zero-momentum mode with single particles. 
In difference to such a homogeneous coherent condensate which corresponds
to a shift of the zero-momentum field operator by a macroscopic c-number,
the homogeneous squeezed condensate (\ref{calE^2sq})
is obtained by the multiplicative operation on the zero momentum modes:
\begin{eqnarray}
U_{\rm sq}^{(0)}{\cal A}_i^a U_{\rm sq}^{(0)-1}&=& 
e^{f_0}{\cal A}_i^a~,\nonumber\\
U_{\rm sq}^{(0)}{\cal E}_i^a U_{\rm sq}^{(0)-1}&=& e^{-f_0}{\cal E}_i^a~.
\label{mult}
\end{eqnarray}
Like the coherent transformation, the squeezing transformation 
is a canonical one, since it leaves the canonical commutator invariant.
Although it is possible in a finite volume 
to relate the squeezed vacuum $|0_{\rm sq}\rangle$
to the perturbative vacuum $|0\rangle$ via the unitary 
squeezing operator, it is important to note that these two vacua become
unitarily inequivalent in the infinite volume limit \cite{miransky}.

\subsection{Gluon selfinteraction and the stability of the squeezed vacuum}
\noindent
After having constructed a squeezed gluon vacuum in an explicit way
we shall see that the gluon selfinteraction is a necessary condition for the
stability of the squeezed vacuum as a candidate for the true QCD vacuum.
For this reason we shall now relate the coefficient $C_0$ to the value of the 
gluon condensate in our model of the squeezed vacuum.
Neglecting the nonzero momentum modes  
we find for the value of the gluon condensate in the squeezed vacuum
\begin{equation}
\langle 0_{\rm sq}|\alpha_s G_{\mu\nu}^a G^{a\mu\nu}|0_{\rm sq}\rangle
_{\rm zero~mom}=
2\langle 0_{\rm sq}|\alpha_s\left({\cal B}^a_i\right)^2|0_{\rm sq}\rangle
-2\langle 0_{\rm sq}|\alpha_s\left({\cal E}^a_i\right)^2|0_{\rm sq}\rangle
\end{equation}
where we have written 
${\cal B}^a_i\equiv g\epsilon_{ijk}f^{abc}{\cal A}_j^b{\cal A}_k^c$.
Noting that the expectation value of the electric field in the squeezed
condensate vanishes, see (\ref{calE^2sq}),
and using Wick's theorem to express the expectation value of the magnetic
field in terms of the contraction $C_0$ we obtain
\begin{equation}
\langle 0_{\rm sq}|\alpha_s G_{\mu\nu}^a G^{a\mu\nu}|0_{\rm sq}
\rangle_{\rm zero~mom}=
2\alpha_s^2{N_c^2-1\over 3 N_c}\left(3N_cC_0\right)^2~.
\label{C_0G^2}
\end{equation}
We see here that  
in our model of a squeezed vacuum of zero momentum gluon pairs 
the self interaction of the gluons is essential for the
coefficient $C_0$ to be nonzero. In an Abelian model $C_0$ would
be zero and the squeezed gluon vacuum constructed in the last 
paragraph unstable. One can regard the
squeezed condensate as a tool for integrating out gluons including
nonperturbative infrared effects.

\section{Effective quark action in the squeezed gluon vacuum}
\label{action}
\noindent
In this section we shall derive the effective quark action of 
Abelian QCD in the presence of the squeezed condensate constructed in the 
last section.

\subsection{Integrating out the gluon fields}
\noindent
We write the generating functional for the massive Abelian QCD model
defined in (\ref{LAbel}) in the form
\begin{eqnarray}
Z_{\rm Abel}[M]&=&\int DA D\Psi D\bar{\Psi}
\exp \left[i\int d^4x{\cal L}_{\rm Abel}[M]\right]\nonumber\\
&\equiv &\int D\Psi D\bar{\Psi}Z[J,M]
\exp \left[i\int d^4x {\cal L}_{\Psi}\right]
\end{eqnarray}
with the generating functional $Z[J,M]$ originating from integrating out 
the gluons.
From the expression (\ref{Hred}) for the reduced Hamiltonian we obtain
in a large but finite volume $V$
\begin{equation}
Z_V[J_{\mu},M]=Z^{(1)}_V[J_0]\cdot Z^{(2)}_V[J_i]
\end{equation}
with the two parts
\begin{eqnarray}
Z^{(1)}_V [J_i,M] &=& 
\langle 0|T\exp\left(i\int d^3x\int dt \tilde{v}^a_i J_i^a\right)|0\rangle~,\\
Z^{(2)}_V[J_0,M] &=& \exp\left[-{i\over 2}\int dt\int d^3x\int d^3y 
J_0^a(\vec{x},t)
{e^{-M|\vec{x}-\vec{y}|}\over 4\pi|\vec{x}-\vec{y}|}J_0^a(\vec{y},t)\right]~.
\label{Zuph}
\end{eqnarray} 
Using Wick's theorem we can write
\begin{equation}
Z^{(1)}_V [J_i,M]= \exp\bigg\{-{1\over 2}\int d^4x\int d^4y J_i^a(x)
\langle 0|T \tilde{v}_i^a(x)\tilde{v}^b_j(y)|0\rangle J_j^b(y)\bigg\}
\label{ZphW}
\end{equation}
with the causal Green function
\begin{eqnarray}
\langle 0|T \tilde{v}^a_i(x)\tilde{v}^b_j(y)|0\rangle &=&
\delta^{ab}\delta_{ij}{{\rm e}^{- iM|x_0-y_0|}\over 2MV}+\nonumber\\
& &\!\! + \delta^{ab}{1\over V}\sum_{q\neq 0}
{\rm e}^{i\vec{q}\cdot (\vec{x}-\vec{y})}
{{\rm e}^{- i \omega (\vec{q})|x_0-y_0|}\over
2\omega(\vec{q})}\left(\delta_{ij}-{q_iq_j\over \vec{q}^2 +M^2}\right)~,
\label{propag}
\end{eqnarray}
where
\begin{equation}
\omega(\vec{q}) = \sqrt{\vec{q}^2+M^2}~.
\end{equation}
With these results for a large finite volume in hand we shall now study
the infinite volume limit $Z[J]\equiv \lim_{V\to\infty} Z_V[J]$ for the two 
different cases, the perturbative and the squeezed vacuum.
For the first case we keep the mass $M$ finite while taking the infinite 
volume 
limit and only afterwards set $M=0$. For the case of the squeezed vacuum
on the other hand we vary the mass $M=1/2C_0V$ with the volume $V$
according to (\ref{msq}).

\subsection{Generating functional for the perturbative vacuum}
\noindent
For the case when $M$ is nonvanishing and constant with respect to $V$, 
we find in the infinite volume limit the standard result
\begin{eqnarray}
\lim_{V\to\infty} \langle 0|T\tilde{v}_i^a(x)\tilde{v}_j^b(y)|0\rangle~ &=&
\delta^{ab}\int {d^3\vec{q}\over (2\pi)^3}
{\rm e}^{i\vec{q}\cdot (\vec{x}-\vec{y})}
{{\rm e}^{- i \omega (\vec{q})|x_0-y_0|}\over
2\omega(\vec{q})}\left(\delta_{ij}-{q_iq_j\over \vec{q}^2 +M^2}\right)
 \nonumber \\ 
&=&\delta^{ab}\int {d^4q\over (2\pi)^4}
i{e^{-iq(x-y)}\over q^2-M^2-i\epsilon}
\left(\delta_{ij}-{q_iq_j\over \vec{q}^2 +M^2}\right)\nonumber\\
&\equiv &\delta^{ab}\tilde{D}_{ij}(M;x-y).
\label{propV}
\end{eqnarray}
Formula (\ref{ZphW}), with (\ref{propV}) inserted, 
together with (\ref{Zuph}) yields
the corresponding generating functional  
\begin{equation}
\label{Z[M]}
Z[J_{\mu},M]=\exp\left\{-{1\over 2}\int d^4x \int d^4y J^{\mu}_a
\tilde{D}_{\mu\nu}(M;x-y)J^{\nu}_a(y) \right\}
\end{equation}
with
\begin{equation}
\tilde{D}_{\mu\nu}(M;x-y)\equiv
\delta_{\mu 0}\delta_{\nu 0}\tilde{D}_{00}(M;x-y)+
\delta_{\mu i}\delta_{\nu j}\tilde{D}_{ij}(M;x-y)~,
\end{equation}
where the spatial components $\tilde{D}_{ij}(M;x-y)$ are given by (\ref{propV})
and the time components are defined by
\begin{equation}
\tilde{D}_{00}(M;x-y)\equiv
-{e^{-M|\vec{x}-\vec{y}|}\over 4\pi|\vec{x}-\vec{y}|}
=\int {d^4q\over (2\pi)^4}
i{e^{-iq\cdot (x-y)}\over \vec{q}^2+M^2}~,
\end{equation}
corresponding to the Yukawa potential in (\ref{Zuph}).
Taking now $M=0$, after the infinite volume limit, and noting that the Abelian
$J^a_\mu$ are conserved, we find the standard generating functional for 
the perturbative gluon vacuum
\begin{equation}
\label{Z_pert}
Z_{\rm pert}[J_{\mu}]=\exp\left\{-{1\over 2}\int d^4x \int d^4y J^{\mu}_a(x)
D(x-y)J^{\mu}_a(y) \right\}
\end{equation}
with the Feynman gauge gluon propagator
\begin{equation}
D(x-y)\equiv -i\int {d^4q\over (2\pi)^4} 
{e^{-iq\cdot (x-y)}\over q^2 +i\epsilon}~.
\label{FGGP}
\end{equation}

\subsection{Generating functional for the squeezed gluon vacuum} 
\noindent
In the case of a squeezed vacuum with $M_{\rm sq}=C_0/(2V)$,  
the two-point Green function (\ref{propag}) becomes in the infinite 
volume limit
\begin{equation}
\lim_{V\to\infty} <0_{\rm sq}|T\tilde{v}^a_i(x)\tilde{v}^b_j(y)|0_{\rm sq}>~=
\delta^{ab}\Big(\tilde{D}_{ij}(M=0;x-y)+ \delta_{ij}C_0\Big)~.
\label{propVsq}
\end{equation}
The extra constant term $C_0$
is due to the presence of the squeezed condensate.
Using (\ref{ZphW}) and (\ref{Zuph}), now with (\ref{propVsq}) instead of
(\ref{propV}), we obtain \footnote{Due to the assumption of global colour neutrality of the 
vacuum in our model, we can make use of the identity 
$1=\exp\big\{-C_0\big[\int d^4x J_0^a(x)\big]^2/2\big\}$ 
in order to write $Z_{\rm sq}[J_\mu]$ in a covariant form.}
\begin{equation}
\label{Z_sq}
Z_{\rm sq}[J_\mu]= Z_{\rm pert}[J_{\mu}]
\cdot\exp\left[-{C_0\over 2}\left(\int d^4x J_\mu^a(x)\right)^2
\right]
\end{equation}
with $Z_{\rm pert}[J_{\mu}]$ given by (\ref{Z_pert}).
In order to obtain the corresponding effective quark action it is useful
to write the squeezed generating functional (\ref{Z_sq}) in the form
\begin{equation}
\label{Z_sq2}
Z_{\rm sq}[J_{\mu}]=\exp\left\{-{1\over 2}\int d^4x \int d^4y J^{\mu}_a(x)
D_{\rm sq}(x-y)J^{\mu}_a(y) \right\}
\end{equation}
with the covariant squeezed propagator
\begin{equation}
D_{\rm sq}(x-y)\equiv D(x-y)+C_0~,
\label{Dsq}
\end{equation}
where $D(x-y)$ is the Feynman gauge gluon propagator (\ref{FGGP}).

\subsection{Effective quark-quark interaction}
\noindent 
From (\ref{Z_sq2}) we obtain the effective quark action
\begin{equation}
W^{\rm eff}_{\rm sq}[\Psi,\bar{\Psi}]=\int d^4x{\cal L}_{\Psi}
+ W_{\rm sq}[\Psi,\bar{\Psi}]~,
\label{Weff}
\end{equation}
where
\begin{equation}
W_{\rm sq}[\Psi,\bar{\Psi}]\equiv -i\ln Z_{\rm sq}=
{1\over 2}i\int d^4x \int d^4y J^{\mu}_a(x) D_{\rm sq}(x-y)J^{\mu}_a(y)~,
\label{LPsieff}
\end{equation}
with $D_{\rm sq}(x-y)\equiv D(x-y)+C_0$. Eq. (\ref{LPsieff}) represents a
modified current-current interaction where in addition to the ordinary bilocal
coupling a permanent and nonlocal coupling of currents with the strength $C_0$
occurs due to the presence of the squeezed condensate.
It represents the zero momentum sector of the theory and is
due to the presence of the squeezed condensate.

Fourier transformation of the four-quark interaction term leads to
\begin{eqnarray}
W_{\rm sq}[\Psi,\overline{\Psi}]
&=&i\frac{g^2}{2}\int {d^4k_1\over (2\pi)^4}...{d^4k_4\over (2\pi)^4}
\overline{\Psi}(k_1){\lambda_a\over 2}\gamma^{\mu}\Psi(k_2)
D_{\rm sq}(k_1-k_2)\nonumber\\
&&\ \ \ \ \ \ \ \ \ \ \ \ \ (2\pi)^4\delta^4(k_1-k_2+k_3-k_4)
\overline{\Psi}(k_3)\gamma_{\mu}{\lambda_a\over 2}\Psi(k_4)~,
\end{eqnarray}
with
\begin{equation}
\label{delta}
D_{\rm sq}(q)= \left[ {i\over q^2+i\epsilon} + (2\pi)^4
\delta^4(q)C_0\right]~.
\end{equation}
The first term corresponds to the usual propagator of a massless boson. It is
responsible for perturbative interactions at large momentum transfer.
The second term is a delta function interaction which may
lead to quark confinement as discussed by Munczek and Nemirovsky \cite{mn}.

We shall now compare our coefficient of the delta function contribution to the 
quark interaction
(\ref{delta}) with that by Munczek and Nemirovsky \cite{mn}, see Eq. 
(\ref{deltaU}). This allows us
to express the coefficient $\eta^2$ in terms of the squeezed condensate
parameter $C_0$ as
\begin{equation}
\label{C0eta} 
\eta^2={16 g^2\over 3}C_0={64\pi\alpha_s\over 3}C_0~.
\end{equation}
Using the relation (\ref{C_0G^2}) which gives the parameter $C_0$
in terms of the gluon condensate $\langle\alpha_s G^2\rangle_{\rm no~quarks}$
in absence of quarks,
\begin{equation}
C_0=\alpha_s^{-1}
\sqrt{{1\over 576\pi}\langle\alpha_s G^2\rangle_{\rm no~quarks}}
\end{equation}
and taking into account the fact that the values of the gluon condensate 
with and without quarks are related via a suppression factor $\gamma$ 
\begin{equation}
\langle\alpha_s G^2\rangle_{\rm phys}=
\gamma \langle\alpha_s G^2\rangle_{\rm no~quarks}~,
\end{equation} 
we obtain the following relation between $\eta^2$ and the physical
gluon condensate 
\begin{equation}
\label{etaG}
\eta^2={8\over 9}
\sqrt{{\pi\over \gamma}\langle\alpha_s G^2\rangle_{\rm phys}}~.
\end{equation}
Using the SVZ value $\alpha_s G^2=0.04~{\rm GeV}^4$ \cite{SVZ} and 
$\gamma= 1/3~(1/2)$ \cite{novikov}
we find the $\eta^2=0.54~(0.44)~{\rm GeV}^2$ respectively.
Recently, Narison \cite{Narison} has ponited out that the SVZ value 
might be too small. In his analysis  of heavy quarkonia
mass splittings in QCD he found an update average value
$\alpha_s G^2=(0.071\pm 0.009)~{\rm GeV}^4$. The corresponding values of 
$\eta^2$ according to (\ref{etaG}) are $\eta^2=0.73~(0.59)~{\rm GeV}^2$ 
for $\gamma =1/3~ (1/2)$ respectively.
Comparing these with the value $\eta^2=1.14~{\rm GeV}^2$ obtained by
Munczek and Nemirovsky \cite{mn} from the fit to experimental
meson spectra, we find rather good agreement. The deviation by a factor
of about two is not too bad in view of the simple models used.

\section{Conclusions}
\label{conclude}
\noindent
The squeezed gluon condensate has become an attractive topic of research
in the last years as an interesting alternative to existing models of the 
QCD gluon vacuum, see e.g. \cite{kogan}.
It is of interest to see which consequences the existence of a squeezed 
gluon condensate has for observable quantities.
One possibility has already been discussed in the earlier work \cite{DB}, 
where it has been found that a squeezed gluon vacuum
can indeed explain the large mass of the $\eta'$ quite successfully. 
In the present work we have considered another consequence, its influence
on the quark-quark interaction.
This would open the possibility to understand the phenomenological
approach by Munczek and Nemirovsky.
Including the gluon selfinteractions in the low energy region 
into an Abelian approximation to QCD in form of a squeezed gluon condensate
we were able to obtain the $\delta$-function type confining part of the
effective gluon two-point function.
Furthermore we were able to estimate the strength $\eta^2$ of this 
interaction from the existing values of the physical gluon condensate.
With the recent estimate of the value of the gluon condensate
 by Narison we have obtained
a value of $\eta^2$ which is only a factor of about two smaller than that
obtained by Munczek and Nemirovsky in their analysis of meson spectra,
which is quite encouraging in view of the simple models used.

We have here only considered a squeezed condensate of pairs of
zero momentum gluons. An interesting extension of the present work would 
be to include nonzero momentum gluon modes also. We would then have to
generalize the contraction $C_0$ to the momentum dependent $C(q)$ 
to include pairs of nonzero relative momentum. 
One could then try to get more detailed
information about the effective confining quark-quark interaction in the QCD
vacuum beyond the momentum $\delta$-function originating from the 
zero momentum gluon pairs in the condensate.
Of course the question of the gauge invariance and Lorentz invariance of the 
corresponding condensate of gluon pairs will be more difficult.
Investigations to obtain a gauge invariant Hamiltonian in the low energy
approximation are in progress \cite{GKMP}.

\nonumsection{Acknowledgements}
\noindent
One of us (V.N.P.) thanks the Max-Planck Gesellschaft for providing a
stipendium during the visit at the MPG AG ''Theoretische Vielteilchenphysik''
Rostock where part of this work has been done.
His work was supported by the Russian Fund for Basic Investigations under
grant No. 96/01-01223 and by the Heisenberg-Landau program.
H.-P. P. acknowledges support by the Deutsche Forschungsgemeinschaft
under grant No. Ro 905/11-2.

\appendix{~Reduced phase space quantization of massive gauge theory}
\subappendix{Reduction of the Lagrangian}
\noindent
In this Appendix we recall the method of reduced phase space quantization
for the case of massive Abelian QCD following Ref.\cite{pav}. This approach
is particularly useful for studies of the massless limit as performed in the
main text.  
The classical action of massive Abelian QCD is
\begin{equation}
W=\int d^4x {\cal L}_(x)=
\int d^4x \left[-{1\over 4}F^{a\mu\nu}F_{\mu\nu}^a+{1\over 2}M^2V_\mu^{a2}+
\bar\Psi(i\not{\!\partial}-m)\Psi-J_a^{\mu}V_{\mu}^a\right] 
\end{equation}
with the Lorentz, but not gauge invariant Lagrangian ${\cal L}_{\rm AQCD}(x)$ 
and the Abelian field strengths 
$F^a_{\mu\nu}=\partial_{\mu}V^a_{\nu}-\partial_{\nu}V^a_{\mu}$.

The Euler-Lagrange equations for $V_0^a$ 
\begin{equation}
{\delta W\over \delta V_0^a}=0 
\end{equation}
are not equations of motion but constraints
\begin{equation}
(\vec{\partial}^2-M^2)V_0^a=-\partial_i\dot{V}_i^a+J_0^a~.
\end{equation}
The fields $V^a_0$ are therefore not dynamical variables and have to be 
eliminated using the constraint equations before
the remaining reduced phase space can be quantized.
The constraint equations correspond in the massless limit to the Gauss laws. 
They can be formally solved for $V_0^a$ by
\begin{equation}
\label{AV_0}
V_0^a[\vec{V},J_0]
={1\over \vec{\partial}^2-M^2}(-\partial_i\dot{V}_i^a +J_0^a)
\end{equation}
and inserted into the original Lagrangian. We have abbreviated
\begin{equation}
{1\over \vec{\partial}^2-M^2} f(\vec{x})=-{1\over 4\pi}\int d^3\vec{y}
{e^{-M|\vec{y}-\vec{x}|}\over |\vec{y}-\vec{x}|}f(\vec{y}).
\end{equation}
The electric field strengths $E_k^a$ then become
\begin{equation}
E_k^a [\vec{V}]= -P_{kj}\dot{V}_j^a
-{1\over \vec{\partial}^2-M^2}\partial_kJ_0^a
\end{equation}
with
\begin{equation}
P_{ij}\equiv
\delta_{ij}-{\partial_i\partial_j\over \vec{\partial}^2-M^2}=
\delta_{ij}^T-{M^2\over \vec{\partial}^2-M^2}\delta_{ij}^{||}~~,
\label{AProjop}
\end{equation}
where we have used the longitudinal and transverse projection operators
$\delta_{ij}^{||}\equiv \partial_i\partial_j/ \vec{\partial}^2$ and
$\delta_{ij}^T\equiv \delta_{ij}-\delta_{ij}^{||}$.
In contrast to the massless case, $P^2\neq P$, 
and $P_{ij}$ is invertible
\begin{equation}
P_{ij}^{-1}=\delta_{ij}^T-{\vec{\partial}^2-M^2\over M^2}\delta_{ij}^{||}=
\delta_{ij}-{\partial_i\partial_j\over M^2}~.
\end{equation}
Inserting this into W we obtain the reduced $W_{\rm red}$
\begin{equation}
W_{\rm red}[\vec{V},\Psi]=\int d^4x {\cal L}_{\rm red}
\equiv \int d^4x \left({\cal L}_{\rm red}^{V} +{\cal L}_{\rm red}^\Psi\right)
\end{equation}
with
\begin{eqnarray}
{\cal L}^V_{\rm red} &=& {1\over 2}\left(\dot{V}_i^a P_{ij}\dot{V}_j^a+
                 V_i^a(\vec{\partial}^2-M^2)P_{ij}V_j^a\right)\nonumber\\
{\cal L}^\Psi_{\rm red} &=& {1\over 2}J_0^a{1\over \vec{\partial}^2-M^2}J_0^a 
      -J_0^a\left({1\over \vec{\partial}^2-M^2}\partial_i\dot{V}_i^a\right)
                  -V_i^aJ_i^a+\bar\Psi(i\not{\!\partial}-m)\Psi~,\nonumber
\end{eqnarray}
with  $P_{ij}$ given by (\ref{Projop}).

We have several choices for the dynamical variables.
Following \cite{pav} we can eliminate the second term in 
${\cal L}_{\rm red}^\Psi$ by introducing the new fermionic variable
\begin{equation}
\label{Apsired}
\tilde{\Psi}\equiv \exp\left(ig{\lambda^a\over 2}{1\over \vec{\partial}^2-M^2}
                         \partial_iV_i^a\right)\Psi~.
\end{equation}
Since (\ref{Apsired}) is only a phase transformation, the corresponding
current $J^a_{\mu}$ stay the same and
${\cal L}^\Psi_{\rm red}$ becomes
\begin{equation}
{\cal L}^\Psi_{\rm red} = {1\over 2}J_0^a{1\over \vec{\partial}^2-M^2}J_0^a
      -J_i^a\left(\delta_{ij}-{\partial_i\partial_j\over \vec{\partial}^2-M^2}
      \right)V_j^a +\bar{\tilde{\Psi}}(i\not{\!\partial}-m)\tilde{\Psi}~.
\end{equation}
In terms of the nonlocal variables 
$\tilde{V}_k^a=P_{kj}V_j^a$ with $P_{ij}$ given by
(\ref{AProjop}) the reduced Lagrangian can be written
\begin{eqnarray}
{\cal L}_{\rm red} &=& {1\over 2}\left(\dot{\tilde{V}}_i^a P^{-1}_{ij}
\dot{\tilde{V}}_j^a+
          \tilde{V}_i^a(\vec{\partial}^2-M^2)P^{-1}_{ij}\tilde{V}_j^a\right)
            \nonumber\\
               & & -J_i^a\tilde{V}_i^a
                   +{1\over 2}J_0^a{1\over \vec{\partial}^2-M^2}J_0^a 
                  +\bar{\tilde{\Psi}}(i\not{\!\partial}-m)\tilde{\Psi}~.
\label{AredL}
\end{eqnarray}
The corresponding canonical conjugate momenta are
\begin{eqnarray}
\tilde{\Pi}_i^a &\equiv & {\delta {\cal L}\over \delta \dot{\tilde{V}}_i^a}
= P^{-1}_{ij}\dot{\tilde{V}}_j^a~,
\nonumber\\ 
\tilde{\Pi} &\equiv & {\delta {\cal L}\over \delta \dot{\tilde{\Psi}}} 
= i\tilde{\Psi}^{+}~.
\end{eqnarray}
The nonlocal variables $\tilde{V}_i^a$ and $\tilde{\Psi}$ 
smoothly turn into the photon field in the massless limit.
The correponding reduced Hamiltonian is found as 
\begin{equation}
H_{\rm red}\equiv H_0+H'
\end{equation}
with the free and interaction parts 
\begin{eqnarray}
H_0 &=& \int d^3\vec{x}\left\{{1\over 2}\left(\tilde{\Pi}_i^aP_{ij}
 \tilde{\Pi}^a_j-\tilde{V}^a_i(\vec{\partial}^2-M^2)P^{-1}_{ij}V^P_j\right)+
\tilde{\Pi}^a\gamma_0(\vec{\gamma}\cdot\vec{\partial}+m)\tilde{\Psi}
\right\}~,\nonumber \\
H' &=&\int d^3\vec{x}
\left(\tilde{V}_iJ_i - {1\over 2}J_0{1\over \vec{\partial}^2-M^2}J_0\right)~.
\nonumber
\end{eqnarray}
Note that the second term in $H'$ corresponds to a Yukawa potential.

\subappendix{Quantization}
\noindent
The theory is now quantized in terms of the nonlocal $\tilde{V}_i^a(x)$
by imposing the canonical commutation relations
\begin{equation}
i[\tilde{\Pi}^a_k(\vec{x},t),\tilde{V}^b_j(\vec{y},t)]
=\delta^{ab}\delta_{kj}\delta(\vec{x}-\vec{y}).
\end{equation}
It has been shown in \cite{pav} that the corresponding Poincar\'e algebra is
fulfilled on both the classical and the quantum level.
The free Hamiltonian is diagonalized by the interaction picture fields
\begin{eqnarray}
\tilde{v}_i^a(\vec{x},t) &=& {1\over V}\sum_{\vec{q}}
{1\over\sqrt{2\omega(\vec{q})}}\sum_{\lambda=1,2,3}
\left(a^a(\lambda,\vec{q})\tilde{\epsilon}_i(\lambda,\vec{q})
e^{-i(\omega(\vec{q})t-\vec{q}\cdot\vec{x})} + h.c.\right)~,\nonumber\\
\tilde{\pi}_i^a(\vec{x},t) &=& -i{1\over V}\sum_{\vec{q}}
\sqrt{{\omega(\vec{q})\over 2}}\sum_{\lambda=1,2,3}
\left(a^a(\lambda,\vec{q})\epsilon_i(\lambda,\vec{q})
e^{-i(\omega(\vec{q})t-\vec{q}\cdot\vec{x})} - h.c.\right)~,\nonumber
\label{Aarepr}
\end{eqnarray}
with the creation and annihilation operators satisfying
\begin{equation}
[a^a(\lambda,\vec{q}),a^{b+}(\lambda',\vec{q}')]=\delta^{ab}
\delta_{\lambda\lambda'}
(2\pi)^3\delta(\vec{q}-\vec{q}')
\end{equation}
and with the real nonlocal polarization vectors 
\begin{equation}
\tilde{\epsilon}_i(\lambda,\vec{q})
\equiv P_{ij}(q)\epsilon_j(\lambda,\vec{q})~.
\end{equation}
Here
\begin{equation}
P_{ij}(q)=\delta_{ij}-{q_i q_j\over \vec{q}^2+M^2}
\end{equation}
is the Fourier transform of the operator (\ref{AProjop})
and $\epsilon_i(\lambda,\vec{q})$ the spatial components of the real covariant
orthonormal polarization vectors $\epsilon_\mu(\lambda,\vec{q})$ satisfying
$\epsilon_\mu(\lambda,\vec{q})\cdot\epsilon_\mu(\lambda',\vec{q})
=\delta_{\lambda\lambda'}$, $q\cdot\epsilon_\mu(\lambda,\vec{q})=0$ and
the completeness relation
$\sum_{\lambda}\epsilon_{\mu}(\lambda,\vec{q})
\epsilon_{\nu}(\lambda,\vec{q})
=-\left(g_{\mu\nu}-q_{\mu}q_{\nu}/ M^2\right)$.
For the nonlocal $\tilde{\epsilon}_i(\lambda,\vec{q})$ we therefore have
the corresponding completeness relation
\begin{equation}
\sum_{\lambda}\tilde{\epsilon}_i(\lambda,\vec{q})
\tilde{\epsilon}_j(\lambda,\vec{q})
=\delta_{ij}-{q_i q_j\over \vec{q}^2+M^2}~.
\end{equation}
This leads to the free propagator
\begin{eqnarray}
\label{Amvecprop}
\langle 0|T\tilde{v}^a_i(\vec{x},x_0)
\tilde{v}^b_j(\vec{y},y_0)|0\rangle &=&
i\delta^{ab}{1\over V}\sum_{\vec{q}}\int {dq_0\over (2\pi)} 
{e^{-iq\cdot (x-y)}\over q^2-M^2 +i\epsilon}
\left(\delta_{ij}-{q_i q_j\over \vec{q}^2+M^2}\right)\nonumber\\
&\equiv&\delta^{ab}\tilde{D}_{ij}(M;x-y)~.
\end{eqnarray}
Combining this with the Yukawa term in $H'$ defining
\begin{equation}
\tilde{D}_{00}(M;x-y)\equiv
i{1\over V}\sum_{\vec{q}}\int {dq_0\over (2\pi)} 
{e^{-iq\cdot (x-y)}\over \vec{q}^2+M^2}~,
\end{equation}
we obtain the complete momentum space propagator
\begin{equation}
\tilde{D}^{\mu\nu}(M;q) = 
{\delta_{\mu 0}\delta_{\nu 0}\over (\vec{q}^2+M^2)}
+{\delta_{\mu i}\delta_{\nu j}\over q^2-M^2+i\epsilon}
\left(\delta_{ij}-{q_iq_j\over (\vec{q}^2+M^2)}\right)~.
\label{Amvecprop2}
\end{equation}
It is the generalization of the photon propagator in Coulomb gauge QED.
In difference to the conventional covariant massive vector propagator 
\begin{equation}
D_{\mu\nu}(M;q)=
-{\delta^{ab}\over q^2-M^2+i\epsilon}\left(g_{\mu\nu}-{q_\mu q_\nu \over M^2}
\right)~.
\label{APhotprop}
\end{equation}
the reduced propagator
(\ref{Amvecprop2}) is regular in the limit $M\rightarrow 0$.
When the vector field is coupled to a conserved fermion current 
$q_{\mu}J^{\nu}=0$ and when no squeezed condensate is present as in QED, 
the two ways of quantization, local and nonlocal, conincide, since 
$J^a_\mu \tilde{D}^{ab}_{\mu\nu}J^b_\nu=J_\mu D_{\mu\nu}J_\nu$.
For the study of the infrared behaviour and in particular in order to construct
a squeezed gluon vacuum it is necessary to have the regular form 
(\ref{Amvecprop2}) of the vector field propagator as discussed in the main 
text. 

\newpage

\subappendix{Lorentz transformation properties}
\noindent
Finally we would like to mention that the following quantum Lorentz 
transformation properties of the field operators $\tilde{v}_k^a$ and 
$\tilde{\pi}_k^a$ have been derived \cite{pav}
\begin{eqnarray}
\delta_L \tilde{v}_k^a &=&
\epsilon_i (x_i\partial_t+t\partial_i)\tilde{v}_k^a - \partial_k 
{1\over \vec{\partial}^2-M^2}\epsilon_i\dot{\tilde{v}}^a_i\\ 
\delta_L \tilde{\pi}_k^a &=&
\epsilon_i (x_i\partial_t+t\partial_i)\tilde{\pi}_k^a 
+\epsilon_i\partial_i(P_{ks}^{-1}\tilde{v}_s^a)
+\epsilon_k\partial_i(P_{is}^{-1}\tilde{v}_s^a)~~.
\end{eqnarray}
This shows that the zero momentum components of the nonlocal fields and their
canonical momenta are invariant under Lorentz boosts.

\nonumsection{References}


\noindent
\begin{thebibliography}{000}
\bibitem{nambu}
 Y. Nambu and G. Jona-Lasinio, Phys. Rev. {\bf 122} (1961) 345. 
\bibitem{njl}
 M.K. Volkov, Ann. Phys. {\bf 157} (1984) 282.\\
 U. Vogl and W. Weise, Prog. Part. Nucl. Phys. {\bf 27} (1991) 195.\\
 S.P. Klevansky, Rev. Mod. Phys. {\bf 64} (1992) 649.\\
 T. Hatsuda and T. Kunihiro, Phys. Rep. {\bf 247} (1994) 221. 
\bibitem{mn}
 H.J. Munczek and A.M. Nemirovsky, Phys. Rev. D {\bf 28} (1983) 181. 
\bibitem{cahill}
 R.T. Cahill, Nucl. Phys. A {\bf 543} (1992) 63c.
\bibitem{rw}
 C. D. Roberts and A. G. Williams, in {\it Progress in Particle and Nuclear
 Physics}, edited by A. Faessler (Pergamon Press, Oxford, 1994), Vol. 33, 477.
\bibitem{tandy}
 P.C. Tandy, in {\it Progress in Particle and Nuclear
 Physics}, edited by A. Faessler (Pergamon Press, Oxford, 1997), Vol. 39, 117.
\bibitem{pag}
 H. Pagels, Phys. Rev. D {\bf 14} (1976) 2747; D {\bf 15} (1977) 2991.
\bibitem{ebert}
 A.A. Bel'kov, D. Ebert and A.V. Emelyanenko, Nucl. Phys. A {\bf 552} 
 (1993) 523. 
\bibitem{fr}
 M. Frank and C.D. Roberts, Phys. Rev. C {\bf 53} (1996) 398. 
\bibitem{BBKR}
 A. Bender, D. Blaschke, Yu. Kalinovsky and C.D. Roberts,
 Phys. Rev. Lett. {\bf 77} (1996) 3724.
\bibitem{bbz}
 M. Baker, J.S. Ball and F. Zachariasen, Nucl. Phys. B {\bf 186} 
 (1981) 531, 560. 
\bibitem{bp}
 N. Brown and M.R. Pennington, Phys. Rev. D {\bf 39} (1989) 2723.
\bibitem{sav}
 G.K. Savvidy, Phys. Lett. B {\bf 71} (1977) 133. 
\bibitem{leut}
 H. Leutwyler, Nucl. Phys. B {\bf 179} (1981) 129. 
\bibitem{celenza}
 L.S. Celenza and C.M. Shakin, Phys. Rev. D {\bf 34} (1986) 1591.
\bibitem{dosch}
 H.G. Dosch, Phys. Lett. B {\bf 190} (1987) 177;\\
 H.G. Dosch and Yu.A. Simonov, Phys. Lett. B {\bf 205} (1988) 339.
\bibitem{biro1}
 T.S. Biro, Ann. Phys. (NY) {\bf 191} (1989) 1;
 Phys. Lett. B {\bf 278} (1992) 15.
\bibitem{mishra1}
 A. Mishra, H. Mishra, S.P. Misra and S.N. Nayak,
 Phys. Rev. D {\bf 44} (1991) 110;
 Z. Phys. C {\bf 57} (1993) 233.
\bibitem{kogan}
 I.I. Kogan and A. Kovner, Phys. Rev. D {\bf 52} (1995) 3719.
\bibitem{DB}
 D. Blaschke, H.-P. Pavel, V.N. Pervushin, G. R\"opke and M.K. Volkov, 
 Phys. Lett. B {\bf 397} (1997) 129;
 {\it Squeezed gluon condensate and the mass of the $\eta'$},
 E-print archive hep-ph/9706528. 
\bibitem{NN}
 N.N. Bogoliubov, J. Phys. {\bf 11} (1947) 23.
\bibitem{pav} 
 H.-P. Pavel and V.N. Pervushin, {\it Reduced phase space quantization of 
 massive vector theory}, E-print archive hep-th/9706220, to appear in 
Int. J. Mod. Phys. A.
\bibitem{stueckel} 
 E.C.G. St\"uckelberg, Helv. Phys. Acta {\bf 30} (1957) 209; see also the 
 discussion in: C. Itzykson and J.-B. Zuber: {\it Quantum Field Theory}
 (McGraw Hill, New York, 1980). 
\bibitem{HeP} 
 W. Heisenberg and W. Pauli, Z. Phys. {\bf 56} (1929) 1;
 {\bf 59} (1930) 160.\\
 P.A.M. Dirac, Can. J. Math. {\bf 2} (1956) 125.\\
 N.S. Han and V.N. Pervushin, Fort. Phys. {\bf 37} (1989) 611.\\
 S.A. Gogilidze, A.M. Khvedelidze and V.N. Pervushin,
 Phys. Rev. D {\bf 53} (1996) 2160.
\bibitem{BSR}
 A. Bender, L. v. Smekal and C.D. Roberts, Phys. Lett. B {\bf 380} (1996) 7.
\bibitem{SVZ} 
 M.A. Shifman, A.I. Vainsthein and V.I. Zakharov, 
 Nucl. Phys. B {\bf 147} (1979) 385.
\bibitem{Narison} 
 S. Narison, Phys. Lett. B {\bf 387} (1996) 162.
\bibitem{miransky}
 For a discussion of the infinite volume limit and unitary inequivalent
 representations, see:
 V.A. Miransky, {\it Dynamical Symmetry Breaking in Quantum Field Theories}
 (World Scientific, Singapore, 1993), Ch. 12.
\bibitem{novikov}
 V.A. Novikov, M.A. Shifman, A.I. Vainshtein and V.I. Zakharov, Nucl. Phys.
 {\bf B 191} (1981) 301.
\bibitem{GKMP}
 S.A. Gogilidze, A.M. Khvedelidze, D.M. Mladenov and H.-P. Pavel,
 Phys. Rev. D {\bf 57} (1998) 7488.
\end{thebibliography}
\end{document}